\documentclass[10pt, conference]{IEEEtran}
\IEEEoverridecommandlockouts

\usepackage{cite}
\usepackage{amsmath,amssymb,amsfonts}
\usepackage{algorithm,algpseudocode}
\usepackage{amsmath}
\usepackage{amsfonts}
\usepackage{amssymb}

\usepackage{graphicx}
\usepackage{textcomp}
\usepackage{xcolor}
\usepackage{comment}
\usepackage{multirow}
\usepackage{booktabs} 
 
\newcommand{\refs}[1]{Section~\ref{#1}}

\newcommand{\reff}[1]{Fig.~\ref{#1}}

\allowdisplaybreaks

\newcommand{\mb}[1]{\mathbf{#1}}

\newcommand{\bseq}{\begin{subequations}}
	\newcommand{\eseq}{\end{subequations}}
\newcommand{\baln}{\begin{align}}
	\newcommand{\ealn}{\end{align}}
\newcommand{\balnd}{\begin{aligned}}
	\newcommand{\ealnd}{\end{aligned}}
\newcommand{\beq}{\begin{equation}}
	\newcommand{\eeq}{\end{equation}}
\newcommand{\beqn}{\begin{eqnarray}}
	\newcommand{\eeqn}{\end{eqnarray}}
\newcommand{\beqno}{\begin{eqnarray*}}
	\newcommand{\eeqno}{\end{eqnarray*}}
\newcommand{\bma}{\begin{displaymath}}
	\newcommand{\ema}{\end{displaymath}}
\newcommand{\bnu}{\begin{enumerate}}
	\newcommand{\enu}{\end{enumerate}}
\newcommand{\bce}{\begin{center}}
	\newcommand{\ece}{\end{center}}
\newcommand{\btb}{\begin{tabular}}
	\newcommand{\etb}{\end{tabular}}
\newcommand{\ba}{\begin{array}}
	\newcommand{\ea}{\end{array}}

\begin{document}
\title{Non-Coherent Massive MIMO Integration in Satellite Communication}
 
\author{\IEEEauthorblockN{\textsuperscript{} Victor Monzon$^*$, Vu N. Ha, Jorge Querol and Symeon Chatzinotas}
\IEEEauthorblockA{\textit{Interdisciplinary Centre for Security Reliability and Trust (SnT),} 
\textit{University of Luxembourg, Luxembourg}\\
\textit{Email:\{victor.monzon, vu-nguyen.ha, jorge.querol, symeon.chatzinotas\}@uni.lu}}
$^*$Corresponding Author: \textit{victor.monzon@uni.lu}
}

\date{}
\maketitle

\begin{abstract}
Massive Multiple Input-Multiple Output (mMIMO) technique has been considered an efficient standard to improve the transmission rate significantly for the following wireless communication systems, such as 5G and beyond. However, implementing this technology has been facing a critical issue of acquiring much channel state information. Primarily, this problem becomes more criticising in the integrated satellite and terrestrial networks (3GPP-Release 15) due to the countable high transmission delay. To deal with this challenging problem, the mMIMO-empowered non-coherent technique can be a promising solution. To our best knowledge, this paper is the first work considering employing the non-coherent mMIMO in satellite communication systems. This work aims to analyse the challenges and opportunities emerging with this integration. Moreover, we identified the issues in this conjunction. The preliminary results presented in this work show that the performance measured in bit error rate (BER) and the number of antennas are not far from that required for terrestrial links. Furthermore, thanks to mMIMO in conjunction with the non-coherent approach, we can work in a low signal-to-noise ratio (SNR) regime, which is an excellent advantage for satellite links.

\end{abstract}

\begin{IEEEkeywords}
Non-coherent, massive MIMO, SatCom
\end{IEEEkeywords}

\section{Introduction}
\label{intro}
 
Nowadays, satellite communication (SatCom) systems have again received a lot of attention thanks to the advances produced in their development technology, which have made it possible to shorten the deployment time of a satellite. The space segment is being revamped, including new technologies such as software-defined radio, new waveforms, and interference management powered by advancing artificial intelligence algorithms that will allow more efficient management of the spectrum of the radio resources that will lead us to have flexible payloads optimized and adapted to a multitude of services tailored according to needs that improve the quality, performance and experience of the user. An extensive review of the advances for the so-called \textit{New Space-Age} is presented in \cite{1}. These advances are leading to a large number of satellite operators planning to launch thousands of non-geostationary (NGSO) satellites to make large constellations of Low Earth Orbit (LEO) and Medium Earth Orbit (MEO) satellites to satisfy the higher demand for satellite-delivered global broadband services and low-latency Internet connections. For instance, the emerging NGSO mega-constellations such as OneWeb, Telesat, and Starlink have a system capacity reaching the terabits-per-second level.

On the other hand, 3GPP began in Release 15 \cite{2} the recommendation to integrate the satellite as one more element of the terrestrial networks. Release 17 \cite{3} completed the standardization considering non-terrestrial networks (NTN) part of the 5G-Beyond standard. This has opened up many lines of interest for research, in which the integration of the satellite poses tremendous challenges due to the difference between the terrestrial and space environments. New radio technologies applied to NTN are studied in \cite{4}, while the insights and challenges are identified in \cite{5}. The integration in 5G and beyond is presented in \cite{6}.

Current wireless networks demand vast amounts of data, which has prompted the search for new, more efficient techniques for the physical layer. Massive MIMO (mMIMO) technology \cite{7,VuHa_TWC18,VuHa_TGCN20}  is considered a primary element of the physical layer due to its energy and spectral efficiency, which consists of using a large number of antennas, mainly on the base station side by the 5G standard. The problem is that this high number of antennas requires the need to estimate a lot of channel information or pilots \cite{8}, leading to long processing times. To solve this problem, non-coherent techniques have been proposed in which it is not necessary to estimate the channel to carry out the detection. These techniques have classically shown a 3~dB degradation in performance versus their coherent counterpart. Non-coherent schemes have been considered for enabling ultra-massive connectivity in sixth-generation networks (6G) \cite{9}. Today, thanks to the conjunction with mMIMO, we can mitigate such degradation by performing the proper system design, as shown in \cite{10} and \cite{11}. The authors in \cite{10} propose a energy-based design for wireless communications.
In contrast, in \cite{11} the authors propose to make the constellation design for the differential phase-based modulation scheme (DPSK). It is verified that for a design based on DPSK, the number of antennas needed for mMIMO is much lower than for one based on energy. In addition, \cite{11} proposes channel coding schemes to reduce by 90$\%$ the number of antennas needed and make it feasible for implementation and deployment. 

Another advantage that phase-based NC-mMIMO schemes have shown is robustness against highly time-dispersive channels with a high Doppler component. The fact that favors its integration in high-speed scenarios, as they showed in \cite{12} and very promising for channels in LEO mega constellations, where the Doppler effect is very present. In the link of a satellite communications system, due to its physical nature, we find a high level of losses and consequences that considerably deteriorate the signal. It is necessary to find modulation and coding schemes or new waveforms that allow these trade-offs to be overcome and make the data link more robust to integrate it with terrestrial networks and meet the requirements imposed by 5G and beyond. Thanks to their versatility, software-defined radios (SDR) are giving rise to new proposals for the components of a SatCom system \cite{11}, where the authors use SDR to cancel interference. All the proposals are based on coherent communication schemes where it is necessary to know and estimate the behaviour of the channel. Errors in this estimate can significantly degrade system performance. 


Therefore, in this paper, we propose to use non-coherent techniques with massive MIMO (NC-mMIMO) to evolve satellite communications systems for the first time in the literature. In this case, due to the need to have a high number of antennas, we analyze where is the best place to allocate NC-mMIMO for different use cases in SatCom, providing the system model in each case, mainly due to these requirements the high number of antennas is assumed by the terrestrial segment, for example, a gateway station. We will present two possible scenarios where we can incorporate the proposed scheme: for geostationary satellites in which the objective is to overcome the high attenuation due to distance and for non-geostationary satellites in low orbit (LEO) whose aim is to overcome the Doppler effect presented by the movement of the constellation. 
  
The rest of the paper is organized as follows. The system model for Non-Coherent massive MIMO schemes based on DPSK applied to SatCom is presented in \refs{sec:model}. The SatCom scenarios are proposed in \refs{sec:sc}. The simulations are presented in \refs{sec:sim}. In \refs{sec:chal} challenges and open issues to integrate NC-mMIMO are identified. Finally, we conclude in \refs{sec:conclusions}.

\section{System Model}
\label{sec:model}

The chosen design to integrate non-coherent schemes in SatCom is based on DPSK due to its robustness in a satellite link. There are two possible configurations in SatCom systems, which are:

\begin{itemize}
    \item Uplink (UL) Scheme: where NC-enabled mMIMO receiver is deployed at the satellite payload.
	\item Downlink (DL) Scheme: where NC-enabled mMIMO receivers are implemented at the ground segment.
\end{itemize}
	
\begin{figure*}[!ht]
\centering
\includegraphics[width=15cm]{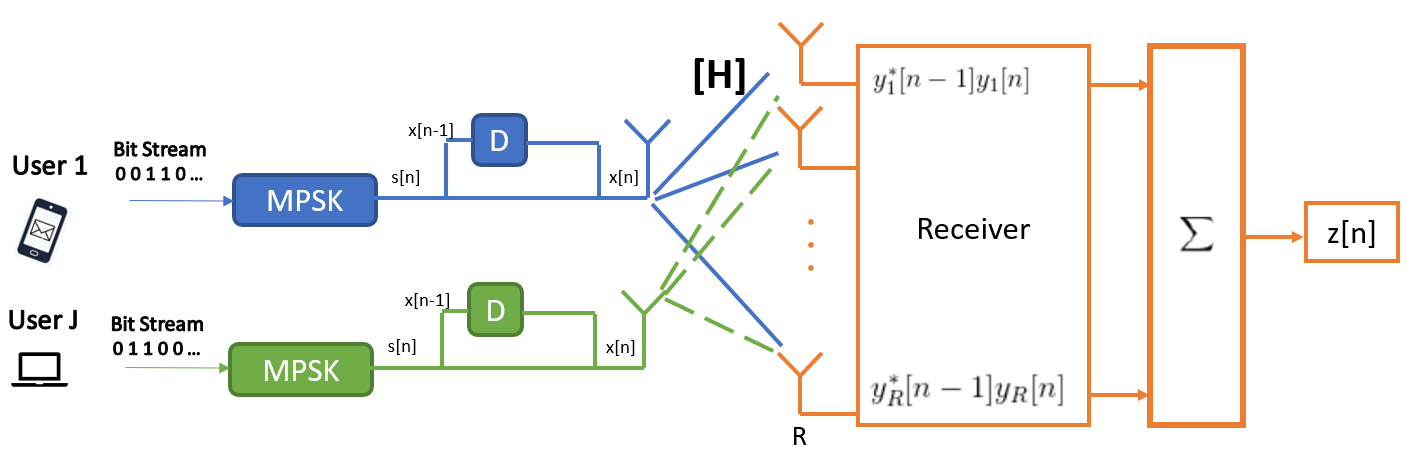}
\caption{System Model for Non-Coherent Communication System.}
\label{Figmodel}
\end{figure*}

The system model for a general non-coherent mMIMO scheme based on DPSK is shown in \reff{Figmodel}. For both schemes, UL and DL, multi-user MIMO scenarios can be considered where there are $R$ receiving antennas equipped at the satellite (UL) or ground station (DL) to communicate to $J$ users (UL) or beams (DL).
Let $x_j[n]$ be the symbol corresponding to user/beam $j$ in time slot $n$. Here, $x_j[n]$ is transmitted by user $j$ and received by satellite in Uplink Scheme, and vice-versa in Downlink Scheme. Following the non-coherent technique, $x_j[n]$, for time index $n>1$, is a differentially encoded version of $s_j [n]$ as \cite{14}. In particularly, 
\begin{equation}
    x_j[n] = x_j[n-1]s_j[n].
\end{equation}

Each $s_j [n]$ symbol belongs to an individual constellation scheme which is different for each user or beam. These constellations are taken from a possible set of base PSK schemes, in which the symbols are distributed along the unit circle according to the criteria established in \cite{11}. Using a constellation by each user provides non-orthogonal multiplexing or access between users. Therefore, depending on the individual constellation chosen, one performance or another will be obtained between users.

The propagation channel is represented by the ($RxJ$)-element channel matrix \textbf{H} with the components $h_{i,j}$ modelling the propagation from user $j$ to the $i-th$ antenna of the receiver (satellite or ground station). For the simulation, we define an example of a satellite link to determine the number of antennas needed in the use cases proposed. On the receiver side (satellite payload in Uplink Scheme or ground users in Downlink Scheme), each antenna receives each time instant signal $y[n]$. Then, all received signals are added to obtain the $z$ variable as shown in \reff{Figmodel}, which can be presented as
\begin{equation}
    z[n]=\dfrac{1}{R}\sum_{i=1}^R y_i^*[n-1]y_i[n]           
\end{equation}

\begin{figure*}[!ht]
\centering
\includegraphics[width=15cm]{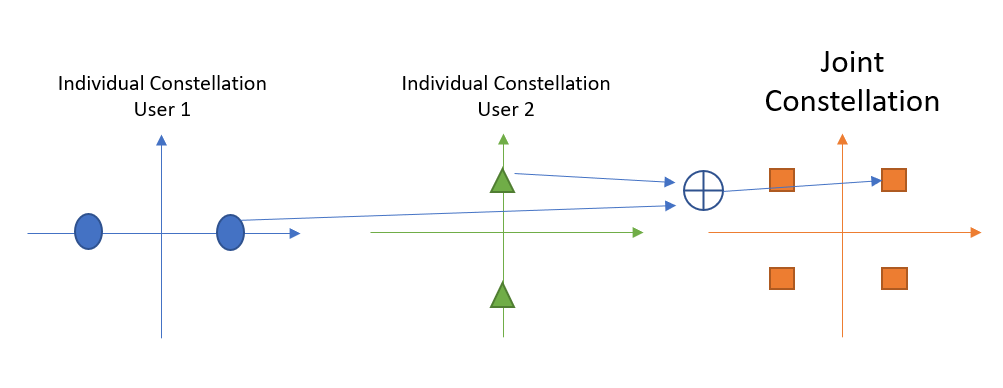}
\caption{Constellation Scheme: Individual (users) and Joint (Receiver).}
\label{constellation}
\end{figure*}

As for the constellation scheme on the receiving side, we have the joint constellation. This constellation is the superposition of all the individual constellations. An example is shown in \reff{constellation}. We have two possible PSK constellations with order $M=2$ (Binary-PSK, BPSK). The individual constellation for user two is a 90-degree rotated version of the constellation for user 1. as shown in \reff{constellation} receiver. This constellation is formed because the channel will non-orthogonally multiplex the two users in the constellation domain. The design of the individual constellations is important since that will depend on the capacity to separate the users in the receiver.

We will refer to each symbol belonging to the joint constellation as joint symbols, represented by the variable $z$ in (2). This value is then used to demultiplexing the individual users based on the knowledge of individual constellation designs. 

\section{Scenarios for Satellite Communications (SatCom)}
\label{sec:sc}

We propose three scenarios for current SatCom that can take advantage of non-coherent schemes empowered by mMIMO for both GEO and LEO satellites.  

\begin{figure}[!ht]
\centering
\includegraphics[width=1\columnwidth]{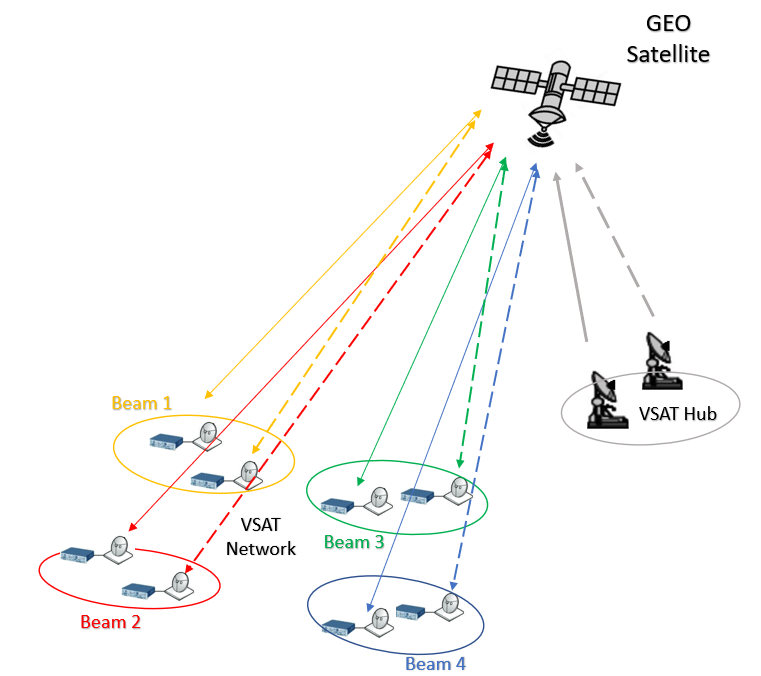}
\caption{Scenario 1: VSAT}
\label{vsat}
\end{figure}

\begin{figure}[!ht]
\centering
\includegraphics[width=1\columnwidth]{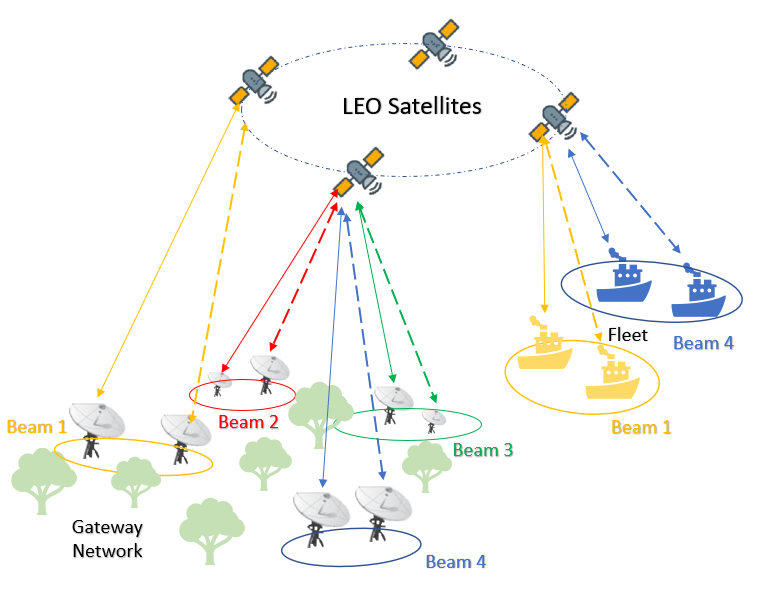}
\caption{Scenario 2: Mega-constellations}
\label{mega}
\end{figure}

\begin{enumerate}
    \item \textbf{VSAT Scenario (UL)}: this scenario includes geostationary satellites (GEO) and very small access terminals (VSAT), as shown in \reff{vsat}. In this context, a non-coherent massive MIMO system is proposed as a multiple access scheme using the constellation domain for the uplink between the VSAT terminals and the satellite. We consider the GEO satellite a multi-beam high throughput satellite (HTS) with terminals covered by the same beam access using different constellation schemes (individual constellations). A group of possible constellations can be defined for each beam. On the other hand, in the uplink between the hub terminal and the GEO satellite, the hubs also access the satellite using different constellation schemes. 
    
    \item \textbf{Mega-constellations Scenario}: the non-coherent communications are used to multiplex different ground segment elements in the emerging mega-constellations formed by low orbit (LEO) or medium orbit (MEO) satellites. In this scenario, as shown in \reff{mega}, we have several cases depending on ground/space assets:
     
    \begin{enumerate}
        \item Gateway stations (GW) on the ground segment access the satellite on the UL, with each GW using a different individual constellation. The set of possible constellations is defined for each beam. The payload uses the joint constellation on-board to demultiplex the users. A GW or teleport station can receive signals from multiple LEO satellites if we consider site diversity in the DL case. In this way, several adjacent satellites can access a GW simultaneously, each using a different individual constellation scheme. It is the GW who uses the joint constellation now in the DL.
        
        \item On the user side, a group of users (\textit{e.g.} fleet of boats) are covered by the same beam and can access the LEO satellite using multiple access in the constellation domain by a non-coherent scheme. In the same way as the GW, a boat can receive signals from different satellites, each of them modulated with a different individual constellation.   
    \end{enumerate}  
    
\end{enumerate}
 
\section{Preliminary Simulation Results}
\label{sec:sim}

For the simulations, we consider an mMIMO system where the \textit{R} receiver antennas are made of a URA antenna array consisting of $R = N_{\sf{ele}}^2$ antennas where $N_{\sf{ele}}$ stands for the number of antenna element per one array side. Furthermore, the ground users are single-antenna equipments. Herein, we modify the channel model presented in \cite{Angeletti_Access20} by regarding Rician model with small-scaled fading parts as in \cite{VuHa_GBC2022} to compare to multipath model used for terrestrial NC-mMIMO in \cite{11,12}.
In particular, the channel coefficient from user $j$ to antenna $i$ of the satellite, $h_{i,j}[t]$ is modeled as,
	\beqn \label{Channel_Model}
	\hspace{-0.5cm} h_{i,j} \!\!\! & = & \!\!\! e^{-j\left(\frac{2\pi d(\theta^{\sf{la}}_j, \theta^{\sf{lo}}_j)}{\lambda}+\phi^i_j\right)} \left[ G^{\sf{gu}}_j / P_{\sf{loss}} \left(\theta^{\sf{la}}_j, \theta^{\sf{lo}}_j \right)\right]^{1/2} \times    \nonumber \\
	\hspace{-0.5cm}&& \hspace{-0.1cm}    \left[ \sqrt{{L}/{(L+1)}}b^{\sf{pa}}_i\left(\theta^{\sf{la}}_j, \theta^{\sf{lo}}_j \right) +  \sqrt{{1}/{(L+1)}}\alpha^i_j \right] , 
	\eeqn
where $G^{\sf{gu}}_j$ is the transmitting antenna gain; 
$P_{\sf{loss}} \left(\theta^{\sf{la}}_j, \theta^{\sf{lo}}_j\right)=\left[ {\lambda}/{4 \pi d(\theta^{\sf{la}}_j, \theta^{\sf{lo}}_j)}\right]^2$ is the path-loss, $\theta^{\sf{la}}_j$  and  $\theta^{\sf{lo}}_j$ are latitude and longitude of user $j$; 
$b^{\sf{pa}}_i\left(\theta^{\sf{la}}_j, \theta^{\sf{lo}}_j \right)$ represents the pattern coefficient of antenna $i$  corresponding to user $j$'s location; $\alpha^i_j$ is the small NLoS fading; $L$ denotes LoS/NLoS Rician factor; $d(\theta^{\sf{la}}_j, \theta^{\sf{lo}}_j)$ is the distance between satellite and user $j$; $\lambda$ is the wave length, and $\phi^i_j$ stands for the phase noise.

In \eqref{Channel_Model}, $b^{\sf{pa}}_i\left(\theta^{\sf{la}}_j, \theta^{\sf{lo}}_j \right)$ and $d(\theta^{\sf{la}}_j, \theta^{\sf{lo}}_j)$ can be determined based on equations $(5)-(7)$ in \cite{Angeletti_Access20}.

\begin{figure}[!ht]
\centering
\includegraphics[width=1\columnwidth]{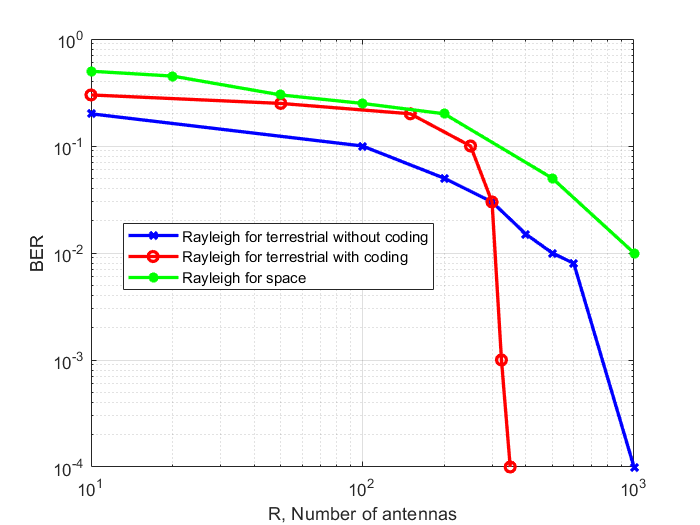}
\caption{Performance for scenario 1}
\label{res1}
\end{figure}

\begin{figure}[!ht]
\centering
\includegraphics[width=1\columnwidth]{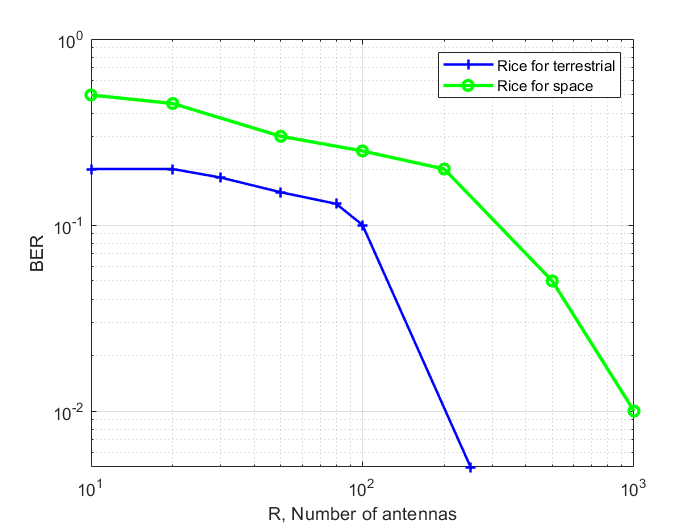}
\caption{Performance for scenario 2}
\label{res2}
\end{figure}

Here, phase noise is one of the imperfections from the hardware components, e.g., oscillators. Then, the channel vector from user $j$ to the satellite can be expressed as
	\beq
	\mb{H}_j = \left[h_{1,j}, h_{2,j},..., h_{R,j} \right].
	\eeq 
Preliminary results for scenario 1 with and without channel coding are shown in \reff{res1} when multipath is present (\textit{L}=0). In this figure, the demodulation of the two possible users in the satellite link when we use the design for non-coherent detection is viable. However, the losses added by the atmosphere have increased the number of antennas, 150 antennas more than in the exclusively terrestrial case. The number of antennas, in this case, is near to without coding case, being slightly increased due to the satellite link conditions. In \reff{res2}, the performance for scenario 2 is shown. In this case, the difference is higher, corresponding to the multi-user interference from DPSK design \cite{13}. However, we can see that the BER is acceptable for 1000 antennas. In this case, it means 700 more antennas than in the terrestrial case. These preliminary results show that non-coherent demodulation for SatCom is possible. The number of antennas has been increased again concerning the terrestrial link as long as we use the same designs as these.

\section{Challenges to Integration in Satellite Communication System}
\label{sec:chal}

There are many challenges to enabling non-coherent techniques for mMIMO SatCom systems. A set of challenges have been identified to be solved from the point of view of the PHY layer using NC waveforms for future payloads. To the best of our knowledge, the challenges identified below have not been considered:
\begin{itemize}
    \item The large propagation delay and the Doppler effects in SatCom may degrade the synchronisation process of the non-coherent scheme; then, the preamble length and periodicity should be re-designed due to these effects.
    
    \item Higher number of users accessing the SatCom systems makes the MPSK modulation design more challenging. Techniques for user scheduling in non-coherent schemes are proposed in [20] based on different fading levels. However, for SatCom is still a challenge considering the high attenuation experienced in the satellite link. Another technique is the combination of non-orthogonal and orthogonal multiplexing with non-coherent detection to increase the system capacity according to the multiple access analysed in \cite{16}.

    \item A limited number of antennas at the payload due to the hardware constraints is also a significant challenge. In the case of the satellite-borne receiver, we must reduce the number of antennas necessary to implement the NC scheme. 
    
    \item The phase shift effect between the antenna elements, usually disregarded in the literature, must be considered in both satellite and ground segments.

    \item The SatCom communications channel differs from the terrestrial networks. The NC designs in previous works have only considered constant channels or non-frequency selective fading models. Hence, understanding the impact of multipath fading channels of SatCom for implementing NC-enabled mMIMO is another critical challenge. In this work, we have simulated a particular multipath channel as an example. However, in dealing with the specific channel model with LoS and NLoS parts in SatCom, one must consider the effect of different fading levels between users. This point would influence the design of the individual constellations discussed in issue 2.
    
    \item For modern HTS, (closed-loop) beamforming or precoding is a widespread technique that requires channel knowledge. For this reason, its evolution to be applied in this type of satellite without knowledge of the channel is quite challenging. In \cite{19} analysis for DL is performed, however, for the scenarios proposed in UL, it has not been considered since the mMIMO should be used in the payload, and the high number of antennas in it is an open issue.
     
    \item The proposed non-coherent empowered by mMIMO schemes are robust to the Doppler effect. For the LEO mega-constellations scenario, a series of channels are studied in the literature and collected in \cite{20}, for which the proposed system must also be validated.
\end{itemize}


\section{Conclusions}
\label{sec:conclusions}

Besides identifying the challenges for implementing NC-based mMIMO in SatCom systems, as presented above, we also aim to analyse problems regarding the difficulties of enabling NC techniques in mMIMO SatCom systems. The analysis answers the challenges posed by creating a feasible NC communications system for the characteristics of future payloads. It thus determines the lines of evolution that the next onboard processors must follow. We extend the channel model to consider different scenarios in SatCom against previous works. Mainly, the results will show the influence of varying channel conditions considering multipath for realistic implementation. Finally, by solving the challenges posed, we can have a non-orthogonal multiple access scheme that provides one more multiplexing domain to improve the management of resources on board the satellite, presenting an alternative to the classic SatCom multiplexing techniques.

\section*{Acknowledgements}
\label{sec:ack}

This work was supported by the FNR project "Dynamic Beam Forming and In-band Signalling for Next Generation Satellite Systems (DISBUS)" under Grant FNR/BRIDGES19/IS/13778945/DISBuS.


\end{document}